\documentclass[
reprint,
superscriptaddress,
showpacs,
nofootinbib,
amsmath,amssymb,
aps,
prl,
floatfix,longbibliography
]{revtex4-2}

\usepackage{graphicx}
\usepackage{dcolumn}
\usepackage{float}
\usepackage{booktabs}
\usepackage[hypertexnames, breaklinks=true, bookmarksnumbered=true, bookmarksopen=true, colorlinks=true, linktocpage=true, citecolor=blue, urlcolor=magenta, linkcolor=magenta]{hyperref}
\usepackage{amsmath,amssymb,amsfonts,amsthm}
\usepackage{mathtools} 
\usepackage[T1]{fontenc}
\usepackage[latin9]{inputenc}
\usepackage{txfonts}
\usepackage{bbm} 
\usepackage{bm} 
\usepackage{mathrsfs} 
\usepackage{xcolor}
\usepackage{natbib}
\usepackage{siunitx}

\newcommand{\ket}[1]{|#1\rangle}

\usepackage{appendix}
\begin{document}

\title{Chiral switching of many-body steady states in a dissipative Rydberg gas}

\author{Chongwu Xie}
\thanks{These authors contributed equally.}
\affiliation{CAS Key Laboratory of Quantum Information, University of Science and Technology of China, \\ Hefei 230026, People's Republic of China}
\affiliation{CAS Center For Excellence in Quantum Information and Quantum Physics, University of Science and Technology of China, Hefei, 230026, People's Republic of China}
\author{Konghao Sun}\thanks{These authors contributed equally.}
\affiliation{CAS Key Laboratory of Quantum Information, University of Science and Technology of China, \\ Hefei 230026, People's Republic of China}
\affiliation{CAS Center For Excellence in Quantum Information and Quantum Physics, University of Science and Technology of China, Hefei, 230026, People's Republic of China}

\author{Kang-Da Wu}\thanks{These authors contributed equally.}
\affiliation{CAS Key Laboratory of Quantum Information, University of Science and Technology of China, \\ Hefei 230026, People's Republic of China}
\affiliation{CAS Center For Excellence in Quantum Information and Quantum Physics, University of Science and Technology of China, Hefei, 230026, People's Republic of China}
\author{Chuan-Feng Li}
\affiliation{CAS Key Laboratory of Quantum Information, University of Science and Technology of China, \\ Hefei 230026, People's Republic of China}
\affiliation{CAS Center For Excellence in Quantum Information and Quantum Physics, University of Science and Technology of China, Hefei, 230026, People's Republic of China}
\affiliation{Hefei National Laboratory, University of Science and Technology of China, Hefei 230088, People's Republic of China}
\author{Guang-Can Guo}
\affiliation{CAS Key Laboratory of Quantum Information, University of Science and Technology of China, \\ Hefei 230026, People's Republic of China}
\affiliation{CAS Center For Excellence in Quantum Information and Quantum Physics, University of Science and Technology of China, Hefei, 230026, People's Republic of China}
\affiliation{Hefei National Laboratory, University of Science and Technology of China, Hefei 230088, People's Republic of China}

\author{Wei Yi}\email{wyiz@ustc.edu.cn}
\affiliation{CAS Key Laboratory of Quantum Information, University of Science and Technology of China, \\ Hefei 230026, People's Republic of China}
\affiliation{CAS Center For Excellence in Quantum Information and Quantum Physics, University of Science and Technology of China, Hefei, 230026, People's Republic of China}
\affiliation{Hefei National Laboratory, University of Science and Technology of China, Hefei 230088, People's Republic of China}

\author{Guo-Yong Xiang}\email{gyxiang@ustc.edu.cn}
\affiliation{CAS Key Laboratory of Quantum Information, University of Science and Technology of China, \\ Hefei 230026, People's Republic of China}
\affiliation{CAS Center For Excellence in Quantum Information and Quantum Physics, University of Science and Technology of China, Hefei, 230026, People's Republic of China}
\affiliation{Hefei National Laboratory, University of Science and Technology of China, Hefei 230088, People's Republic of China}
\begin{abstract}
Dissipative Rydberg gases are an outstanding platform for the investigation of many-body quantum open systems.
Despite the wealth of existing studies, the non-equilibrium dynamics of dissipative Rydberg gases are rarely examined or harnessed from the perspective of non-Hermitian physics, which is but intrinsic to open systems.
Here we report the experimental observation of a chiral switching between many-body steady states in a dissipative thermal Rydberg vapor, where the interplay of many-body effects and non-Hermiticity plays a key role.
Specifically, as the parameters are adiabatically varied around a closed contour, depending on the chirality of the parameter modulation, the Rydberg vapor can change between two collective steady states with distinct Rydberg excitations and optical transmissions.
Adopting a mean-field description, we reveal that both the existence of the bistable steady states and chiral dynamics derive from an exceptional structure in the parameter space, where multiple steady states of the many-body Liouvillian superoperator coalesce.
We demonstrate that both the exceptional structure and the resulting state-switching dynamics are tunable through microwave dressing and temperature variations,
confirming their reliance on the many-body dissipative nature of the Rydberg vapor.
\end{abstract}

\maketitle

\section{Introduction}
Many-body quantum open systems host rich collective behaviors that are absent in isolated settings.
Understanding and manipulating these often complicated phenomena have become an active field of research in recent years, thanks to the rapid progress in engineering quantum simulators across a range of physical systems including ultracold atoms, superconducting qubits, trapped ions, and Rydberg gases~\cite{qsrev}.
A quantum open system is intrinsically non-Hermitian~\cite{nhrev}, and subject to effective non-Hermitian descriptions, in contrast to their Hermitian counterparts of isolated (or closed) systems.
As such, many intriguing properties of non-Hermitian Hamiltonians also manifest in the dynamics of quantum open systems, offering unique perspectives and fresh insights.

An outstanding example here is the branching singularity in non-Hermitian models known as the exceptional point (EP), where eigenvalues and eigenvectors of the non-Hermitian Hamiltonian simultaneously coalesce~\cite{ep1966,ep1998,ep1999,ep1,ep2,eptopo,ep3,ep4,ep5}.
Intriguingly, these Hamiltonian EPs give rise to critical behaviors and topological phenomena, which have stimulated extensive studies on EP-enhanced sensing~\cite{sen1,sen2,sen3}, laser-mode selection~\cite{lms1,lms2,lms3}, directional wave-packet transport~\cite{dwpt1}, and chiral state transfer~\cite{ec2,ec3,ec5,ec6,ec7,cla1,cla2,phot1,sq1,sq2,ca,nv,ions}.
In quantum open systems where the density-matrix dynamics is driven by the Liouvillian superoperator, EPs can also emerge in the Liouvillian eigenspectrum (dubbed Liouvillian EPs)~\cite{LEP1,LEP2,LEP3}, and are responsible for the recently observed chiral-state transfer in transient dynamics~\cite{sq1,sq2} and efficient quantum heat engines~\cite{qhe}.
However, the experimental studies of the Liouvillian-EP-related dynamics are limited to the single-particle case, as an experimentally accessible quantum many-body system that features EP-induced dynamics remains elusive.
Further, since the Liouvillian EPs typically occur in the excited eigenstates (away from the low-lying steady states with vanishing Liouvillian eigenvalues), their impact is often limited to transient dynamics, which presents a practical challenge for experimental observation.


\begin{figure*}[tbp]
   \centering
	\includegraphics[width=0.8\textwidth]{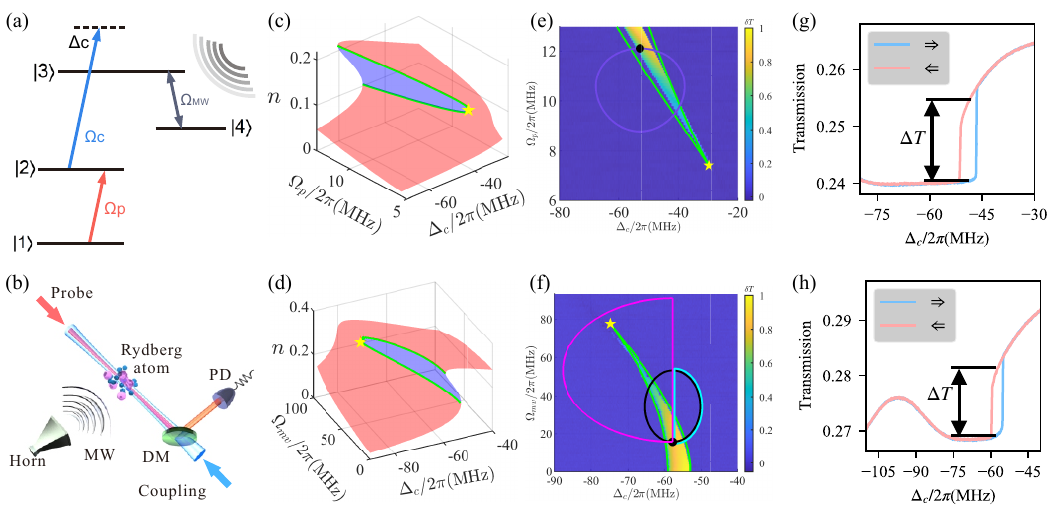}
	\caption{\label{fig:phase_bis}
		\textbf{Exceptional structures in a thermal Rydberg gas}.
  (a) Schematics of the coupling scheme, with a microwave dressing. Labeling of the energy levels: $\ket{1}\equiv\ket{5S_{1/2}}$, $\ket{2}\equiv\ket{5P_{3/2}}$, $\ket{3}\equiv\ket{50D_{5/2}}$, $\ket{4}\equiv\ket{51P_{3/2}}$. The parameters $\Omega_c$, $\Omega_p$, and $\Omega_{\text{mw}}$ represent the Rabi frequencies of the corresponding couplings, respectively.
    (b) Experimental setup, with detailed parameters given in the Methods.
    The parameters of lasers and microwave in this experiment can be found in Appendix.
  Meaning of the abbreviations: PD, Photoelectric detector; DM, dichroic mirror; MW, microwave.
   (c) Numerically calculated total Rydberg-state population $n=\rho_{33}+\rho_{44}$ of the steady states as a function of $\Omega_p$ and $\Delta_c$, without the  microwave field.
   (d)  Numerically calculated total Rydberg-state population $n$ as a function of $\Omega_{mw}$ and $\Delta_c$ with microwave field.
   (e)(f) Experimentally measured phase diagrams under the parameters of (c)(d), respectively.
      The color bar represents the normalized transmission difference, $\delta T$, between the forward and backward scans of $\Delta_c$.
   The green lines represent the numerically fitted second-order ELs, and the yellow star is the third order EP.
The parameters are: (c)(e) $\Omega_c/2\pi=\SI{12.21}{\mega\hertz}$. (d)(f) $\Omega_p/2\pi=\SI{13.07}{\mega\hertz}$, $\Omega_c/2\pi=\SI{12.21}{\mega\hertz}$. See Methods for fitting parameters used in the theoretical model.
(g)(h) Sample data for the extraction of $\Delta T$, showing hysteresis behavior in the presence of bistable steady states. The red and blue arrows indicate the scanning direction, and $\Delta T$ is the transmission difference between two opposite scanning directions.
We take $\Omega_p/2\pi=11.55~\text{MHz}$ and $\Omega_{mw}/2\pi=0$ in (g), and $\Omega_p/2\pi=13.07~\text{MHz}$ and $\Omega_{mw}/2\pi=15.67~\text{MHz}$ in (h).
   The small peak in (h) at $\Delta_c/2\pi\approx -100$ MHz corresponds to the EIT of the state $|50D_{3/2}\rangle$.
  }
\end{figure*}

\begin{figure*}[tbp]
   \centering
	\includegraphics[width=0.8\textwidth]{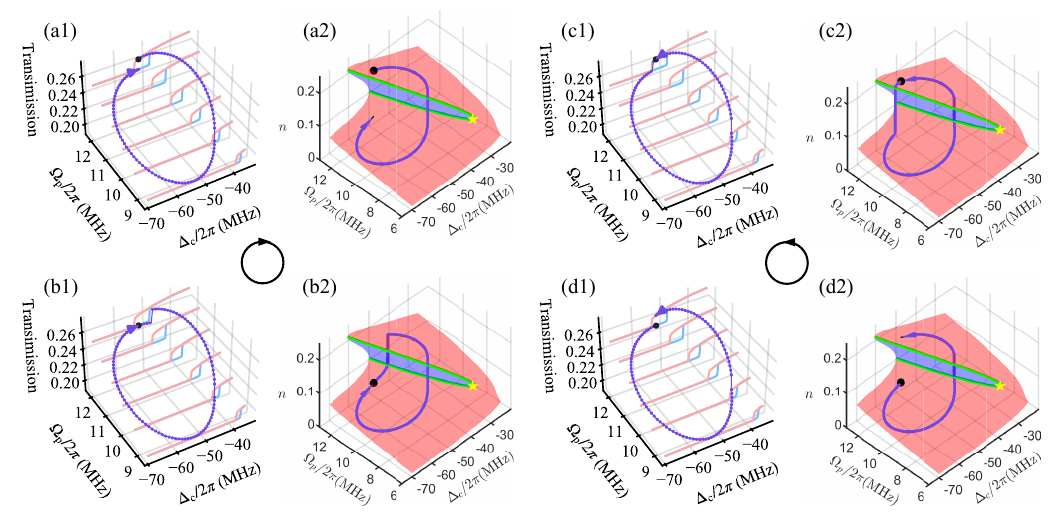}
	\caption{\label{fig:encircle_light}
		\textbf{Chiral mode switching without the microwave field.}
   (a1)(c1)/(b1)(d1) The purple dots, connected by the purple lines to guide the eye, represent the measured transmission along the purple trajectory in Fig.~\ref{fig:phase_bis}(e). We show samples of the measured hysteresis of the transmission in the background, to better visualize the steady states.
   Here, the black dots represent the initial states, and the arrows indicate the encircling  direction.
   (a2)-(d2) The corresponding numerically calculated encircling trajectories on the landscape of the steady-state Rydberg population. The parameters are the same as those in Fig.~\ref{fig:phase_bis}(c).
  }
\end{figure*}

In this work, we experimentally demonstrate adiabatic chiral state switching in a dissipative Rydberg vapor, where, under the many-body effects therein, Liouvillian EPs emerge in the degenerate steady-state subspace and impact the long-time dynamics.
Characterized by high-lying electronic states and strong interactions, Rydberg atoms are an ideal candidate for simulating many-body phenomena such as quantum magnetism~\cite{qm1,qm2,qm3,qm4}, topological order~\cite{topo1,topo2,topo3}, as well as many-body dynamics and thermalization~\cite{ned1,ned2,ned3,ned4,ned5}.
Dissipative Rydberg gases, particularly thermal Rydberg vapors, further provide direct access to non-equilibrium dynamics in many-body quantum open systems~\cite{carr_2013_prl,trv1,trv2,trv3,trv4}.




Our experiment is based on the bistable transition in a thermal Rydberg vapor, where two collective steady states feature different Rydberg excitations, and are characterized by distinct light transmissions in the standard electromagnetically induced transparency (EIT).
While the density-matrix dynamics of the system can be captured by a many-body Lindblad master equation under the mean-field approximation, an exceptional structure emerges in the parameter space where degenerate steady states of the Liouvillian superoperator coalesce.
Ramification of the steady states gives rise to second-order exceptional lines (ELs) in the parameter space that terminate at a third-order EP.
We experimentally demonstrate that the dynamics near the exceptional structure exhibits chiral features, exemplified by a chiral state switching between different steady states---the light transmission can be switched to a different value when the parameters are adiabatically tuned around a closed contour to come back to the starting point, but the switching only occurs  one-way around.
Such a chiral state switching is the many-body counterpart of the EP encircling observed in single-particle non-Hermitian models~\cite{ec3,ec5}. It sensitively depends on the many-body parameters of the Rydberg vapor, and is tunable through microwave dressing and temperature modulation.
Further, since both the chiral dynamics and the exceptional structure vanish on the single-atom level, our observation is fundamentally different from the previously studied chiral dynamics in classical non-linear non-Hermitian systems~\cite{nlep2,nlep4} where the emergence of exceptional structures in the non-linear case hinges upon the presence of Hamiltonian EPs without non-linearity.
Our findings introduce a fresh non-Hermitian perspective to the dissipative dynamics in Rydberg atoms, with interesting prospects for future applications.

\begin{figure*}[tbp]
   \centering
	\includegraphics[width=0.8\textwidth]{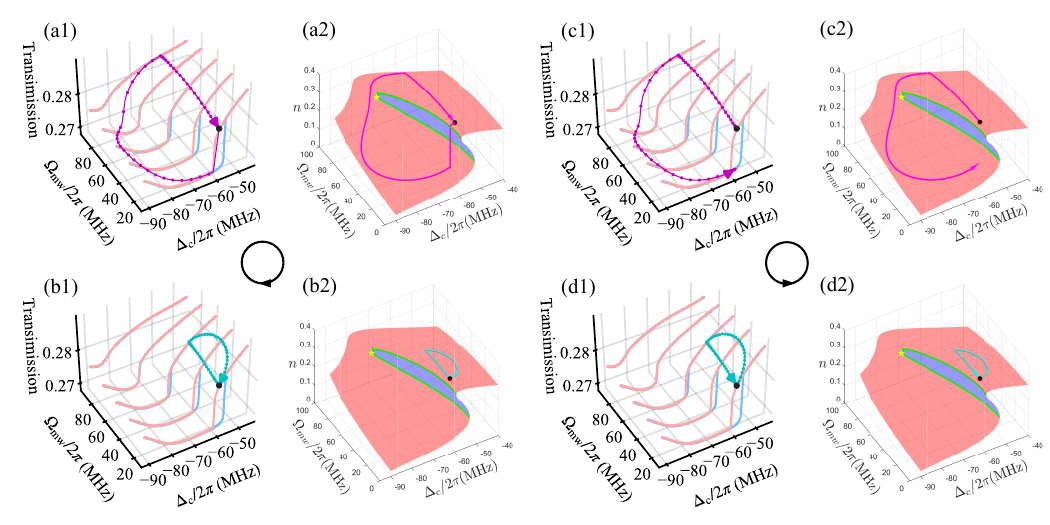}
	\caption{\label{fig:encirlc_mw}
		\textbf{Chiral mode switching with the microwave-field dressing.}
  (a1)(c1)/(b1)(d1) The cyan (magenta) dots, connected by the cyan (magenta) lines to guide the eye, represent the measured transmission along the cyan (magenta) trajectory in Fig.~\ref{fig:phase_bis}(f). We show samples of the measured transmission hysteresis in the background.
  The black dots represent the initial states, and the arrows indicate the encircling direction, respectively.
  (a2)-(d2) The corresponding numerically calculated encircling trajectories on the landscape of the steady-state Rydberg population. The parameters are the same as those in Fig.~\ref{fig:phase_bis}(d).
}
\end{figure*}

\begin{figure*}[tbp]
   \centering
	\includegraphics[width=0.8\textwidth]{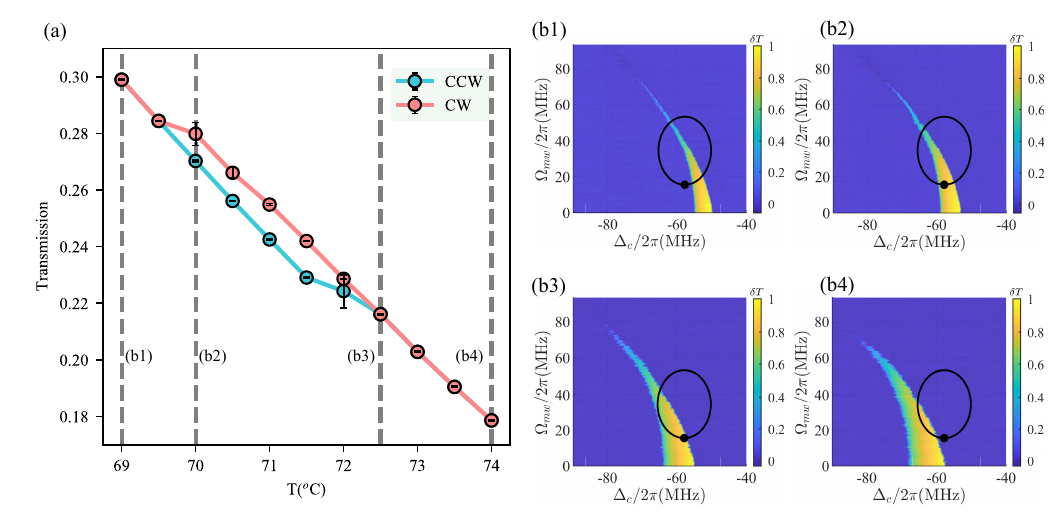}
	\caption{\label{fig:tem}
		\textbf{Tuning Chiral mode switching using temperature}.
  (a) The transmission of the probe laser  at the end of the trajectory, as a function of the temperature.
  The red (cyan) line represents measurements from the clockwise (counter-clockwise) encircling direction.
  The error bars are the standard deviation for 50 repeated measurements.
  (b1)-(b4) The relative location of the exceptional structure to the fixed trajectory (black loop), as the temperature changes: (b1) 69 $^o$C, (b2) 70 $^o$C, (b3) 72.5 $^o$C and (b4) 74 $^o$C.
  }
\end{figure*}

\section{Experimental setup and theoretical model}
For our experiment, we use a $^{87}$Rb vapor cell at a temperature of
$\sim 70^\circ$C. As illustrated in Fig.~\ref{fig:phase_bis}(b),
the cell is illuminated by a pair of counter propagating laser beams: a weak probe light with a central wavelength of 780 nm (resonant with the transition $5S_{1/2}\Longleftrightarrow5P_{3/2}$) and a typical power of several micro Watts;
and a strong coupling light with a central wavelength of 480 nm (blue-detuned to the transition $5P_{3/2}\Longleftrightarrow 50D_{5/2}$). An additional microwave field, with a central frequency of 17.047 GHz and the same polarization as that of the incident light, is applied to the atoms to resonantly couple the Rydberg states $50D_{5/2}$ and $51P_{3/2}$.
The coupling scheme is illustrated in Fig.~\ref{fig:phase_bis}(a),
where the relevant internal states are labeled $\ket{1}\equiv\ket{5S_{1/2}}$, $\ket{2}\equiv\ket{5P_{3/2}}$, $\ket{3}\equiv\ket{50D_{5/2}}$, $\ket{4}\equiv\ket{51P_{3/2}}$ throughout the work.

In the absence of the microwave field, the atoms undergo the standard EIT in a ladder-type configuration. Under the many-body effects of the Rydberg gas, the system is known to host
two collective steady states with distinct occupations in the Rydberg state $\ket{3}$. The addition of the microwave field further dresses the Rydberg states, affecting the population of the Rydberg excitations, as well as the properties of the steady states.
Invoking a mean-field treatment of the many-body effects~\cite{mw_nc}, the dynamics of the atomic ensemble is described by the Lindblad equation
\begin{align}
    \dot{\rho}=\mathcal{L}\rho:=-i[H,\rho]+\sum_{i=1}^6 \left( L_i \rho  L_i^{\dagger} -\frac{1}{2} \{ L_i^\dagger L_i,\rho\} \right)\label{eq1},
\end{align}
where $\mathcal{L}$ is the Liouvillian superoperator, and
\begin{align}
    H=\begin{pmatrix}
     0 & \Omega_p/2 & 0 & 0 \\
      \Omega^{*}_p/2 & 0 & \Omega_c/2 & 0\\
        0 & \Omega_c^{*}/2  & -\Tilde{\Delta}_1 & \Omega_{mw}/2\\
        0 & 0 & \Omega_{mw}^{*}/2 & -\Tilde{\Delta}_2
      \end{pmatrix},
\end{align}
in the basis $\{|1\rangle, |2\rangle, |3\rangle, |4\rangle\}$.
The dissipative processes are described by the quantum jump operators $L_i$, with $L_1=\gamma_{21}| 1 \rangle\langle 2 |$, $L_2=\gamma_{32}| 2 \rangle\langle 3|$, $L_3=\gamma_{31}| 1 \rangle\langle 3|$, and $L_5=\gamma_{41}| 1 \rangle\langle 4|$
accounting for the corresponding spontaneous emissions, respectively.
The dephasing of the Rydberg states are given by $L_4=\gamma_3| 3 \rangle\langle 3 |$ and $L_6=\gamma_4| 4 \rangle\langle 4|$.
The effective detuning $\Tilde{\Delta}_1=\Delta_c-\alpha\rho_{33}-\beta\rho_{44}$ and $\Tilde{\Delta}_2=-\gamma\rho_{33}-\epsilon\rho_{44}$ incorporate the many-body effects in the Rydberg gas on the mean-field level, which are proportional to the population of the Rydberg states. The many-body effects can derive from the long-range Rydberg interactions~\cite{carr_2013_prl,trv2,mw_nc} or ion-induced
Stark shifts~\cite{plasma1,plasma2}.
Here $\gamma_{21}, \gamma_{31}, \gamma_{32}, \gamma_{41}$ and $\gamma_{3},\gamma_4$ are the corresponding spontaneous decay rate and dephasing rate, respectively, and $\{\alpha,\beta,\gamma,\epsilon\}$ are the many-body parameters.
Note that in the absence of the microwave field, the Rydberg state $|4\rangle$ is decoupled, and one should neglect jump operators $L_{5,6}$, focusing only on the dynamics within the subspace spanned by $\{|1\rangle,|2\rangle,|3\rangle\}$.

Setting $\dot{\rho}=0$ in Eq.~(\ref{eq1}) allows one to solve for the steady states of the system, which correspond to the eigenstates of the Liouvillian superoperator with vanishing eigenvalues.
Under the influence of the many-body effects, the steady state can be degenerate, characterized by multiple solutions of the Rydberg-state population under the same parameters.
In Fig.~\ref{fig:phase_bis}(c)(d), we show the steady-state populations of the Rydberg states $n =\rho_{33}+\rho_{44}$ (shaded in red), both without and with the microwave field. In either case, the steady state is three-fold degenerate within a parameter window (shaded in blue). Further analysis shows that only the highest- and lowest-lying solutions are stable, indicating bistability.
Nevertheless, at the boundaries of the bistable region (green curves), a stable steady state coalesces with the unstable one. The boundaries are hence identified as the second-order Liouvillian ELs, which further terminate at a third-order EP (yellow pentagrams), where all three steady-state solutions merge.

Note that the steady-state degeneracy and the Liouvillian exceptional structures derive from the many-body effects of the Rydberg gas, as they disappear for vanishing many-body parameters $\{\alpha,\beta,\gamma,\epsilon\}$.
The exceptional structure here is therefore fundamentally different from that in classical non-linear non-Hermitian models~\cite{nlep2,nlep4,nlep1,nlep3}, where EPs already emerge in the eigenspectrum of the linear Hamiltonian and become modified under non-linearity. Furthermore, despite the emergence of the exceptional structure in the steady-state space, all steady states possess vanishing Liouvillian eigenvalues, in contrast to the branching eigenvalues at the Hamiltonian EPs. The dynamic impact of the exceptional structures therefore manifests at long times, when the system is in the vicinity of the steady state. This contrasts previous studies of Liouvillian EPs in the excited eigenstates, which focuses on the dynamics at short times.

Experimentally, the bistable region outlined by the exceptional structure is observable through the transmission of the probe laser in the EIT.
Specifically, with other parameters fixed, we scan the detuning $\Delta_c$ forth and back, and record the difference in the transmission $\Delta T$.
In Fig.~\ref{fig:phase_bis}(e)(f), we show the experimental data for the normalized transmission difference $\delta T = \Delta T /\Delta T_{max}$ between the forward and backward scans, in the planes of $\Omega_p-\Delta_c$ and $\Omega_{mw}-\Delta_c$, respectively.
Here $\Delta T_{max}$ is the maximum value of $\Delta T$ throughout the scanned parameter space.
The bistable region is clearly visible in Fig.~\ref{fig:phase_bis}(e)(f), with non-vanishing $\delta T$.
Since $\Delta T$ is positively correlated to the difference of the Rydberg-state population $n$ between the two steady states, the boundary of the bistable region corresponds to that of the Liouvillian exceptional structure, which agrees well with the numerical fit (green lines) using Eq.~(\ref{eq1}).
In Fig.~\ref{fig:phase_bis}(g)(h), we plot the measured transmission data under fixed $\Omega_p$ and $\Omega_{mw}$. The observed hysteresis is nothing but the well-known collective bistable transition, between a steady state with a large Rydberg excitation and transmission, and another with limited excitation and transmission~\cite{carr_2013_prl}.
Importantly, now that we identify the bistable transition points with the Liouvillian ELs and EPs, a natural question arises as to their dynamic consequences.

\section{Chiral mode switching across the EL}
A notable dynamic consequence of EPs in non-Hermitian models is the chiral state transfer.
As the system parameters slowly vary along a closed loop in the vicinity of an EP, the final state of the encircling dynamics depends on the chirality of the parameter change. The phenomenon is a direct result of the branching eigenspectral landscape of the non-Hermitian Hamiltonian near the EP, and has been observed in classical systems with gain and loss~\cite{cla1,cla2}, and in
quantum open settings where the dynamics is dominated by single qubits~\cite{ec2,phot1,sq1,sq2,ions,ca,ec7}.
Here we show that chiral mode switching persists in the many-body setting of Rydberg vapor, thanks to the similarity of the collective steady-state landscape near the Liouvillian exceptional structure to that of the eigenspectrum near a Hamiltonian EP.

For our experiment, we initialize the system in one of the steady states within the bistable region, and slowly vary $\Omega_p$ and $\Delta_c$ along the purple loop in Fig.~\ref{fig:phase_bis}(e). Each encirclement takes the time $T=40$ s, and for each data point, we stay the system for $100$ ms before taking data, to ensure its relaxation to the local steady state. Both timescales are much larger than the relaxation time $T_r\approx 0.2$ ms near the bistable boundary  (see Supplementary Material for detailed protocol).
Figure~\ref{fig:encircle_light}(a1)(c1) show the measured transmission when the initial state is the high-lying steady state, and in the absence of microwave dressing. Under a clockwise (CW) parameter change, the measured transmission changes to a lower value on returning to the initial point, indicating a state switch. By contrast, under a counterclockwise (CCW) rotation, the measured transmission recovers its initial value, as the system returns to its initial steady state. The chirality of the state switching is reversed when the system is initialized in the low-lying steady state, as illustrated in Fig.~\ref{fig:encircle_light}(b1)(d1). For each case, we show the corresponding Liouvillian steady-state landscape to the right [see Fig.~\ref{fig:encircle_light}(a2)(b2)(c2)(d2)], where the purple trajectory corresponds to the numerically simulated population using Eq.~(\ref{eq1}) with the encircling period $T=5$ ms. The experimental observations are in good agreement with numerical simulations. Note that the trajectory of the encirclement intersects the second-order EL on either
side of the third order EP at least once, which, as we show
later, is important for the occurrence of chiral state switching.

\section{Tuning the many-body effects}
Since the observed chiral state switching is many-body in nature, it is also subject to controls that impact the many-body effects in the Rydberg vapor. Indeed, according to Eq.~(\ref{eq1}) and Fig.~\ref{fig:phase_bis}(f), through a resonant coupling of two Rydberg states, the microwave dressing significantly changes the bistable region and the exceptional structure. Further, since the power of the microwave field can be tuned over a wider range than that of the probe field, the chiral dynamics can now be examined over a broader range of parameters.

With the microwave dressing, we experimentally demonstrate the chiral state switching in the parameter space of $\Omega_{mw}$ and $\Delta_c$.
We vary the parameters along the closed trajectories shown in Fig.~\ref{fig:phase_bis}(d), all with an encircling period of $T=20$ s.
Facilitated by the improved tunability of $\Omega_{mw}$, we initialize the system in a high-lying steady state within the bistable region, but choose different encircling trajectories.
As illustrated in Fig.~\ref{fig:encirlc_mw}(a1)(c1), the chiral state switching persists when the trajectory cuts through the bistable region [magenta in Fig.~\ref{fig:phase_bis}(d)], thus intersecting the EL on either side of the third-order EP once. This is similar to the trajectory in Fig.~\ref{fig:encircle_light}.
However, when the trajectory only intersects with the EL on one side of the EP [green in Fig.~\ref{fig:phase_bis}(d)], the measured transmission always recovers the initial value, as the system returns to the same steady state. This is illustrated in Fig.~\ref{fig:encirlc_mw}(b1)(d1).
Based on these observations, we summarize a set of sufficient conditions for the occurrence of chiral state switching:
i) the initial state is prepared in the bistable region as one of the steady states;
ii) with respect to the initial position, the two nearest intersection points between the trajectory and the ELs should be on either side of the third order EP.

Alternatively, the atomic density of the vapor is tunable through temperature, which offers a novel control over the chiral state switching.
To demonstrate such a possibility, we fix the encircling contour in the parameter space of  $\Omega_{mw}$ and $\Delta_c$, as well as the initial state which is chosen
as the high-lying steady state.
We then perform CW and CCW encirclings over a fixed trajectory at different temperatures. Since the exceptional structure shifts toward regions with larger detuning $|\Delta_c|$ [see Fig.~\ref{fig:tem}(b1-b4)], varying the temperature adjusts the relative position of the fixed trajectory to the exceptional structure, enabling the tuning of the chiral state switching.
We now speed up the encircling process and vary the parameters around the loop within $T=33$ ms.
Since this time scale is still much larger than the local relaxation time, we expect the system to remain close to the local steady states along the trajectory.
We then measure the  transmission of the final state of the CW and CCW rotations. As illustrated in Fig.~\ref{fig:tem}(a), the chiral state switching emerges in the intermediate temperature range, where the two conditions for the chiral state switching hold.


\section{Summary}
We experimentally demonstrate the chiral state switching between many-body steady states in a dissipative Rydberg vapor.
Through theoretical analysis, we reveal that the phenomenon, along with the well-known bistable transition in Rydberg gases, is a direct consequence of the Liouvillian exceptional structure, consisting of ELs and higher-order EPs.
We establish the condition for the occurrence of the chiral state switching through experiments
with various parameters and encircling trajectories.
Importantly, both the exceptional structure and the chiral dynamics studied here hinge upon the many-body nature of the Rydberg gas, and are therefore fundamentally different from previous studies on two- or three-level systems~\cite{sq1,sq2}.
Such a many-body origin not only provides novel control schemes in the form of the microwave dressing and temperature modulation, but also offers a practical route toward the rich non-Hermitian physics in the quantum many-body regime.
For instance, various extensively researched properties of EPs in non-Hermitian settings, the EP-enhanced sensing in particular, may find applications in the context of many-body quantum open systems that are experimentally accessible.



\textbf{Acknowledgement.} This work was supported by the National Natural Science Foundation of China (Grants Nos. 12134014, 61905234, 11974335, 12374337, and 12374479), the Key Research Program of Frontier Sciences, CAS (Grant No. QYZDYSSW-SLH003), USTC Research Funds of the Double First-Class Initiative (Grant No. YD2030002007) and the Fundamental Research Funds for the Central Universities (Grant No. WK2470000035).



\clearpage
\appendix

\pagebreak
\widetext
\begin{center}
\textbf{\large  Supplemental Material for "Chiral switching of many-body steady states in a dissipative Rydberg gas"}
\end{center}
\setcounter{equation}{0}
\setcounter{figure}{0}
\setcounter{table}{0}
\makeatletter
\renewcommand{\theequation}{S\arabic{equation}}
\renewcommand{\thefigure}{S\arabic{figure}}
\renewcommand{\bibnumfmt}[1]{[S#1]}
\renewcommand{\citenumfont}[1]{S#1}

\section{Theoretical fitting}
For the boundary of the bistable region in the absence of the microwave field, we fix $\Omega_c/2\pi=\SI{12.21}{\mega\hertz}$ according to the experiment, and vary $\gamma_{21}$, $\gamma_{32}$, $\gamma_{31}$, $\gamma_{3}$, and $\alpha$ to fit the calculated ELs with the measured boundaries of the bistable region. The fitted values of the parameters are: $\gamma_{21}/2\pi=\SI{27.22}{\mega\hertz}$, $\gamma_{32}/2\pi=\SI{0.29}{\mega\hertz}$, $\gamma_{31}/2\pi=\SI{0.016}{\mega\hertz}$, $\gamma_3/2\pi=\SI{20.31}{\mega\hertz}$ and $\alpha/2\pi=\SI{-273.89}{\mega\hertz}$.

In the presence of the microwave field, we fix $\Omega_c/2\pi=\SI{12.21}{\mega\hertz}$, $\Omega_p/2\pi=\SI{13.07}{\mega\hertz}$ according to the experiment, and vary $\gamma_{21}$, $\gamma_{32}$, $\gamma_{31}$, $\gamma_{3}$, $\alpha$, $\beta$, $\gamma$, and $\epsilon$ to fit the calculated ELs with the measured boundaries. We further require $\gamma_{41}=\gamma_{31}$ and $\gamma_3=\gamma_4$ for the fitting process. The fitted parameters are: $\gamma_{21}/2\pi=\SI{7.34}{\mega\hertz}$, $\gamma_{32}/2\pi=\SI{0.70}{\mega\hertz}$, $\gamma_{31}/2\pi=\gamma_{41}/2\pi=\SI{0.13}{\mega\hertz}$, $\gamma_3/2\pi=\gamma_4/2\pi=\SI{11.78}{\mega\hertz}$, $\alpha/2\pi=\SI{-174.32}{\mega\hertz}$, $\beta/2\pi=\SI{-954.93}{\mega\hertz}$, $\epsilon/2\pi=\SI{1114.08}{\mega\hertz}$, and $\gamma/2\pi=\SI{6843.66}{\mega\hertz}$.

\textbf{Stability of the steady state}
To study the stability of the steady-state solution, we consider a small deviation from
the steady-state solution $\rho'$, with $\rho=\rho'+\Delta\rho$. Here $\rho'$ satisfies $\mathcal{L}[\rho']=0$.
We then expand Eq.~(\ref{eq1}) to the first order in $\Delta\rho$ around $\rho'$. For instance,
in the absence of the microwave field, we have
\begin{align}
	\Delta\dot{\rho}= J\left[\rho^\prime\right]\Delta\rho,
\end{align}
where
\begin{widetext}
	\begin{align}
		&\resizebox{\textwidth}{!}{$
			J\left[\rho^\prime\right] =
			\begin{pmatrix}
				0 & 0 & -\Omega_p & 0 & 0 & \gamma_{21} & 0 & 0 & \gamma_{31} \\
				0 & -\frac{\gamma_{21}}{2} & 0 & 0 & -\frac{\Omega_c}{2} & 0 & 0 & 0 & 0 \\
				\frac{\Omega_p}{2} & 0 & -\frac{\gamma_{21}}{2} & \frac{\Omega_c}{2} & 0 & -\frac{\Omega_p}{2} & 0 & 0 & 0 \\
				0 & 0 & -\frac{\Omega_c}{2} & \frac{1}{2} (-\gamma_{3}-\gamma_{32}-\gamma_{31}) & \Delta_c-\alpha \rho^{\prime}_{33} & 0 & 0 & \frac{\Omega_p}{2} & -\alpha \text{Im} \rho^{\prime}_{13}\\
				0 & \frac{\Omega_c}{2} & 0 & \alpha \rho^{\prime}_{33}-\Delta_c & \frac{1}{2} (-\gamma_{3}-\gamma_{32}-\gamma_{31}) & 0 & -\frac{\Omega_{p}}{2} & 0 & \alpha \text{Re} \rho^{\prime}_{13} \\
				0 & 0 & \Omega_p & 0 & 0 & -\gamma_{21} & 0 & -\Omega_c & \gamma_{32} \\
				0 & 0 & 0 & 0 & \frac{\Omega_p}{2} & 0 & \frac{1}{2} (-\gamma_{21}-\gamma_{3}-\gamma_{32}-\gamma_{31}) & \Delta_c-\alpha \rho^{\prime}_{33} & -\alpha \text{Im} \rho^{\prime}_{23} \\
				0 & 0 & 0 & -\frac{\Omega_p}{2} & 0 & \frac{\Omega_c}{2} & \alpha \rho ^{\prime}_{33}-\Delta_c & \frac{1}{2} (-\gamma_{21}-\gamma_{3}-\gamma_{32}-\gamma_{31}) & \frac{1}{2} \left(2 \alpha \text{Re} \rho^{\prime}_{23}-\Omega_c\right) \\
				0 & 0 & 0 & 0 & 0 & 0 & 0 & \Omega_c & -\gamma_{32}-\gamma_{31} \\
			\end{pmatrix}$},\\
		&\Delta\rho =\left[\Delta\rho_{11},\mathrm{Re}\Delta\rho_{12},\mathrm{Im}\Delta\rho_{12}, \mathrm{Re}\Delta\rho_{13},\mathrm{Im}\Delta\rho_{13}, \Delta\rho_{22}, \mathrm{Re}\Delta\rho_{23},\mathrm{Im}\Delta\rho_{23}, \Delta\rho_{33}\right]^\text{T}.
	\end{align}
\end{widetext}
We now examine the eigenvalues of the non-Hermitian matrix: when the real components of the eigenvalues are all negative, the steady state $\rho'$ is stable; otherwise, $\rho'$ is unstable.
The stable (unstable) regimes in Figs.~\ref{fig:phase_bis}(b) and are thus determined.
In the presence of the microwave field, we follow a similar procedure to calculate the stable (unstable) regimes in Fig.~\ref{fig:phase_bis}(d).

\begin{table}
	\centering
	
	\begin{tabular}{cccc}
		\toprule
		Figures & $\Omega_p/2\pi$ (MHz)& $\Omega_{mw}/2\pi$ (MHz)& Temperature$~(^oC)$ \\
		\midrule
		Figure~\ref{fig:encircle_light} & 5.51-12.96 & 0 & 70
		\\
		Figure~\ref{fig:encirlc_mw} & 13.07   & 0-94.02 & 70
		\\
		Figure~\ref{fig:tem} & 13.07   & 0-94.02 & 69.0-74.0
		\\
		\bottomrule
	\end{tabular}
	
	\caption{\label{tab:parameters} \textbf{Experimental parameters.}
		Both the microwave and probe laser are resonant to the corresponding energy levels in the whole experiment.
		The power of the coupling laser is fixed at about 127.3 mW which corresponds to a Rabbi frequency $\Omega_c/2\pi=12.21$ MHz.
		Both the two lasers are focused to the same location and have a same $1/e^2$ waist of around 100 $\mu$m.
		All the lasers and microwave have the same polarization.
	}
\end{table}

\section{Experimental parameters}
Table~\ref{tab:parameters} shows the general experimental parameters.
In the demonstration of the chiral dynamics in the Rydberg system, as  shown in Figs.~\ref{fig:encircle_light} and \ref{fig:encirlc_mw}, we change the parameters every time interval $\tau= 0.2$ s.
At each time interval, we wait $0.1$ s for the system to evolve to a steady state and record the transmission of the probe laser with an integral time of $0.1$ s.
The discrete temporal parameters along the purple loop in Fig.~\ref{fig:phase_bis}(e) are as follows
\begin{align}
	\Omega_p(n)&= \Omega_{0}\sqrt{1 + p \cos{\theta_n}},\
	\Delta_c(n)=\Delta_0 + \delta\sin{\theta_n},
\end{align}
where $\Omega_{0}/2\pi=10.58$ MHz, $p=0.31$, $\Delta_0/2\pi=-52.80$ MHz, $\delta/2\pi = 14.92$ MHz.
$\Omega_p(n)$ and $\Delta_c(n)$ are the parameters at the time $t_n=n\tau$.
The value of $\theta_n$ depends on the encirclement direction: $\theta_n=2\pi n/N$ for the clockwise and $\theta_n=2\pi (1-n/N)$ for the counterclockwise directions, respectively.
$n$ is taken from 0 to $N$, and $n=0$ is the starting while $n=N$ is the ending.
Here, we set $N=200$, which decides the time cost in an encirclement.
On the other hand, the discrete temporal parameters along the three loops in Fig.~\ref{fig:phase_bis}(f) are given by
\begin{align}\label{eq:loopmw}
	\Omega_{mw}(n)&=\Omega_0 - \Tilde\Omega\cos{\theta_n},\ \Delta_c(n) = \Delta_0 - \delta \sin{\theta_n},
\end{align}
with different $\Omega_0,\ \Tilde\Omega,\ \Delta_0$ and $\delta$, which can be found in Tbl.~\ref{tab:loops}.
As before, we set $\theta_n=2\pi n/N$ for the clockwise and $\theta_n=2\pi (1-n/N)$ for the counterclockwise directions, respectively, with $N=100$.
\begin{table}
	\centering
	\begin{tabular}{ccccc}
		\toprule
		loops & $\Omega_0/2\pi$ (MHz)& $\Tilde\Omega/2\pi$ (MHz)& $\Delta_0/2\pi$ (MHz)& $\delta/2\pi$ (MHz) \\
		\midrule
		black & 34.48 & 18.81  & -57.80 & 7.54
		\\
		cyan & 34.48  & 18.81  & -57.80 & 7.54 $\Theta(\pi-\theta_n)$
		\\
		magenta & 53.27  & 37.60 & -57.80 & 29.86 $\Theta(\theta_n-\pi)$
		\\
		\bottomrule
	\end{tabular}
	
	\caption{\label{tab:loops} \textbf{Description of loops.}
		Here $\Theta(x)=1$ when $x>0$ and $\Theta(x)=0$ when $x\leq0$.
		These parameters are used in Eq.~\eqref{eq:loopmw}
	}
\end{table}

\section{Experimental details}
\begin{figure*}
   \centering
	\includegraphics[width=0.8\textwidth]{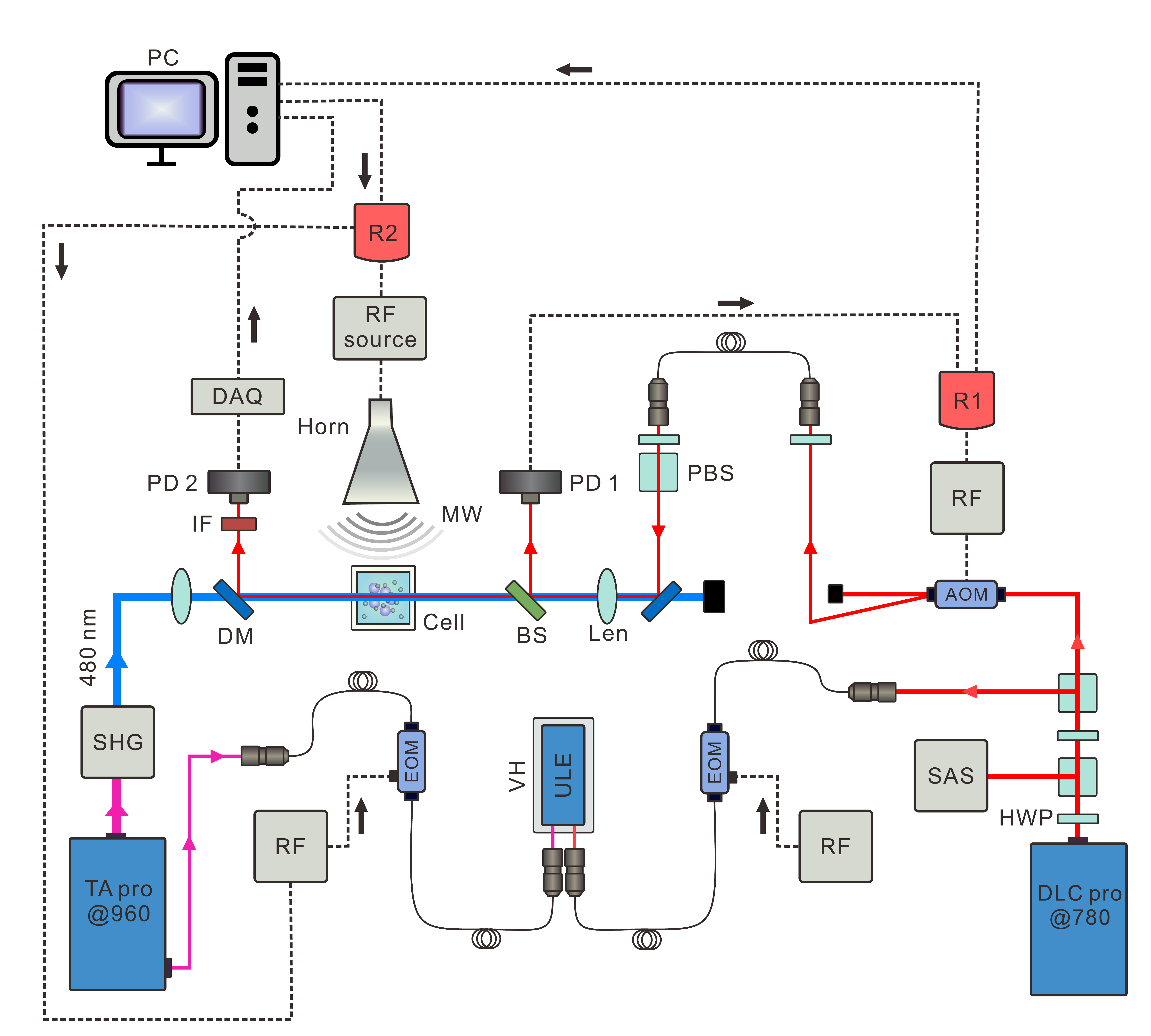}
	\caption{\label{fig:detailedexpsetup}
		\textbf{Detailed experimental setup.} The whole experimental setup can enable us to experimentally control several different parameters, e.g., laser detuning, laser power, MW power. Meaning of the abbreviations: R, RedPitaya; PD, Photoelectric detector; DM, dichroic mirror; MW, microwave; IF, interference filter; RF, radio frequency source; EOM, electro-optical modulator; AOM, acousto-optical modulator; PBS, polarizing beam splitter; BS; beam splitter; SHG, second harmonic generation; ULE, ultra-low expansion materials; SAS, saturation absorption spectrum.}
\end{figure*}
\textbf{Parameter control}.
The detailed information of our experimental setup for investigating parameter encircling is detailed in Fig.~\ref{fig:detailedexpsetup}.
In our experiment, $\Omega_p,\ \Delta_c,$ and $\Omega_{mw}$ are the three control parameters.
First, a photoelectric detector (PD, PD1 in Fig.~\ref{fig:detailedexpsetup}) is used to monitor the power of the laser. An acousto-optical modulator (AOM) and a RedPitaya-designed servo system (R1 in Fig.~\ref{fig:detailedexpsetup}) help lock the power of the probe laser $\Omega_p$ to a fixed value.
Redpitaya can function as a lockbox by utilizing the pyrpl Python module.
We can easily modify the set point of the servo system to vary the value of $\Omega_p$ continually using Python codes.
Second, the frequency of the coupling laser is locked on a cavity.
Actually, the coupling laser is generated through a second harmonic generation (SHG) process using a basic laser with a central wavelength of 960 nm.
We split a little basic laser (about 10 mW) to a frequency lock system before the SHG process (see Fig.~\ref{fig:detailedexpsetup}).
The split basic laser is modulated through an electro-optic modulator (EOM) before it enters the cavity.
The EOM not only makes a modulation to generate an error signal but can also make a large frequency shift on the basic laser.
That means we can easily modify the frequency of the basic laser used in the SHG by changing the frequency of the radio frequency (RF) signal on the EOM.
Then the coupling laser detuning $\Delta_c$ can be controlled.
However, if we directly switch the frequency of the RF by setting another value, the RF will output nothing in its switching time.
And the frequency of the coupling laser cannot change continually using such a method.
So we apply the frequency modulation (FM) function to the RF source to control the frequency of the RF, and $\Delta_c$ continually.
The FM function can make a frequency shift proportional to the modulation voltage, and a continually varying voltage can be easily obtained experimentally using an arbitrary signal generator (ASG).
RedPitaya can work as an ASG, which can be easily controlled with Python.
So we can use a computer to control the RedPitaya (R2 in Fig.~\ref{fig:detailedexpsetup}), and subsequently $\Delta_c$ with the help of the FM.
For the same reason as that of the microwave source, we can apply an amplitude modulation (AM) to the microwave source to control $\Omega_{mw}$ continually.

\textbf{Generation of the probe and coupling lasers}.
A continues laser (Toptica DLC pro) with a central wavelength of 780 nm and a total power of 80 mW is split into three branches: the first is used for monitoring the absolute frequency by saturation absorption spectrum (SAS) technology, the second is sent to frequency locking to $5S_{1/2}\leftrightarrow5P_{3/2}$ transition and linewidth reduction using a high-finesse reference cavity, and the third branch is used for experiments, i.e., the probe light in the EIT system. Moreover, the power of the probe light is monitored by a photoelectric detector (PD1) and stabilized using an AOM.

The coupling light, which has a central wavelength of 480 nm (resonant with $5P_{3/2}\leftrightarrow50D_{5/2}$), is generated through a SHG step with Toptica tapered amplifier (TA) pro. In particular, the TA unit can deliver up to 1.4 W at 960 nm and is then fed to the frequency conversion module using a periodically poled potassium titanyl phosphate (PPKTP) crystal, which is anti-reflecting coated at both 480 nm and 960 nm, and the temperature is stabilized to around 28$^\circ$C to achieve the quasi-phase match condition. As a result, for an input power of 1.385 W, an output power of 200 mW of the coupling laser is achieved. Note that we have sacrificed  conversion efficiency to achieve a more stable performance.

\textbf{Linewidth reduction and frequency control of the probe and coupling lasers}.
For reducing technical noise and conveniently controlling the frequency of both probe and coupling lasers, we use a double-coated 100-mm-long high finesse-reference cavity made of ultra-low expansion (ULE) material, which is loaded on a Zerodur mounting block with four contacting viton balls for vibration isolation. The cavity is placed in a vacuum house (VH, model VH-6020-4), which is enclosed by loose thermal insulation, with the temperature stabilized to around 32.06 $^\circ$C. The typical vacuum degree is on the order of 10$^{-8}$ mbar after baking. The finesses of 780 nm and 960 nm light are around 15000 and 20000, respectively. After locking the frequency to the cavity mode with a fast PID (FALC 110) using PDH technology, the linewidth is supposed to be around 1 kHz. The whole frequency locking system, except the locking electronics, is provided by a stable laser system (SLS).

Since the cavity mode cannot be coincident with both the probe and coupling laser mode, we let both 960 and 780 nm light be coupled into a fiber electro-optical modulator for generating sidebands with a large frequency range (9 kHz-10 GHz) and locked one of the sidebands of both lasers to one cavity mode. Thus, the frequency of both the probe and coupling lasers can be arbitrarily tuned by controlling the RF source.

\begin{figure}
   \centering
	\includegraphics[width=0.45\textwidth]{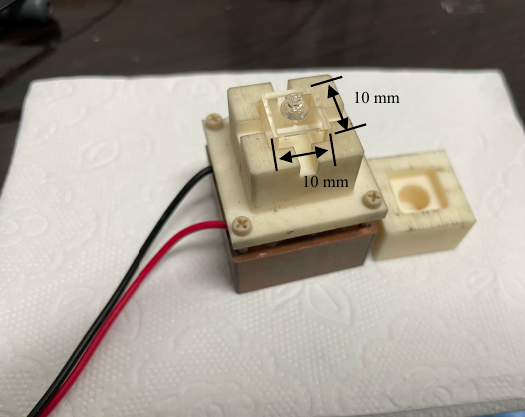}
	\caption{\label{fig:cell}
		\textbf{Vapor cell and ceramic oven.}
  The vapor cell is placed in a ceramic oven.}
\end{figure}

\textbf{Vapor cell and temperature control}.
The vapor cell used in this experiment has a tiny internal geometric size of $8\times 8\times 8~\text{mm}^3$.
It is filled with natural rubidium vapor ($27.8\%$ for $^{87}$Rb and $72.2\%$ for $^{85}$Rb), but we only used the part of $^{87}$Rb.
The vapor cell is contained in a specially designed ceramic oven (see Fig.~\ref{fig:cell}).
Ceramics have good thermal conductivity and do not reflect or absorb microwaves like metals do.
The oven is driven by a thermoelectric cooler (TEC), which is controlled by a TEC controller (TED4015, Thorlabs).

\section{Additional data}

\begin{figure*}
   \centering
	\includegraphics[width=1\textwidth]{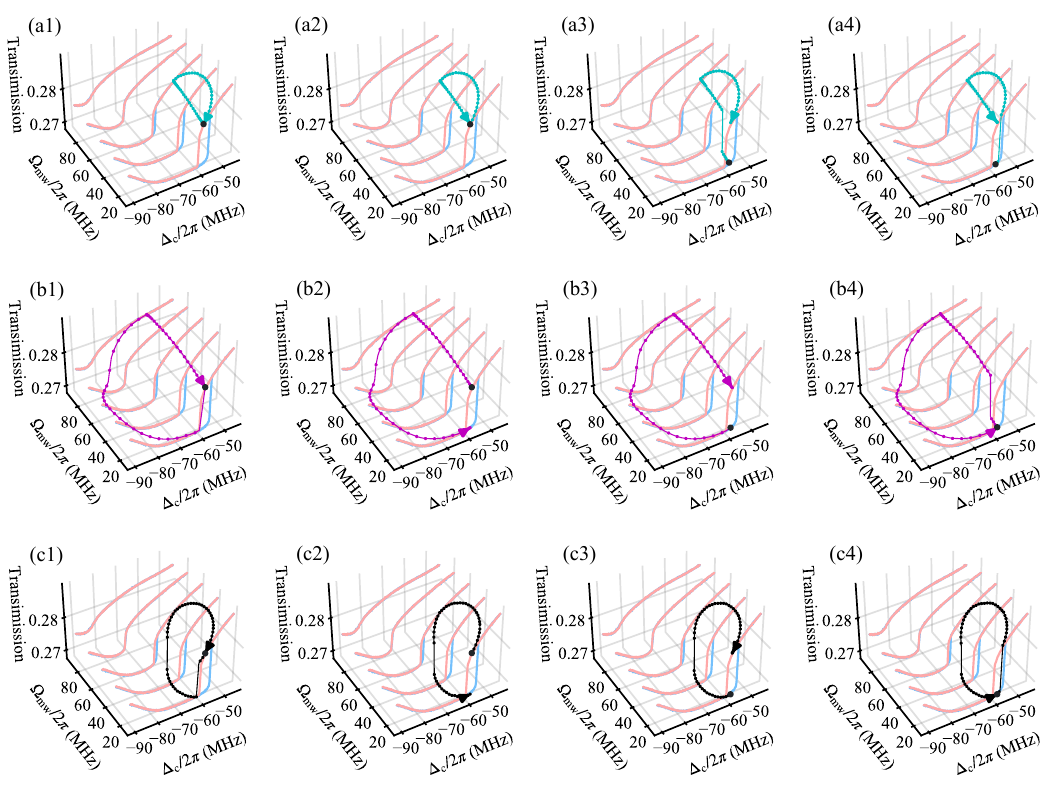}
	\caption{\label{fig:additional}
		\textbf{Additional encircling.} The encircling trajectories are along the path in Fig.~\ref{fig:phase_bis}(f) with the same color. The meanings of the points, lines and arrows are the same as those in Figs.~\ref{fig:encircle_light} and ~\ref{fig:encirlc_mw}.
  }
\end{figure*}

\textbf{Additional encircling paths.}
We have only displayed a portion of the encirclement's results due to the main text's space constraints.
In fact, we have achieved the encirclement along a variety of loops experimentally; they will be shown below as supplemental material.
Figure~\ref{fig:additional} shows the more detailed encircling results along the loops in Fig.~\ref{fig:phase_bis}(f). Figure~\ref{fig:additional}(a1)-(a4) do not show a chiral phenomenon where the final states are independent on the encircling direction or the initial states.
As expected, Figs.~\ref{fig:additional}(b1)-(c4) show chiral switching where the final states are only dependent on the encircling direction.
Since they satisfy the two conditions for the occurrence of chiral switching.

\begin{figure}
   \centering
	\includegraphics[width=0.45\textwidth]{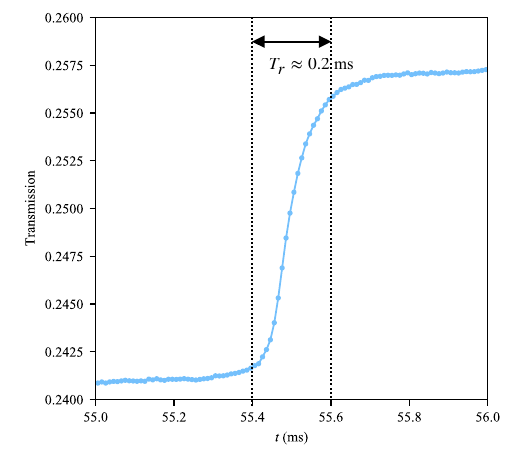}
	\caption{\label{fig:ev}
		\textbf{State evolution.}
  The measured transmission of the probe laser when we scan $\Delta_c$. Here the parameters are the same as those in Fig.~\ref{fig:phase_bis}(g). }
\end{figure}

\textbf{Relaxation time near the bistable region.} When we scan $\Delta_c$ across the bistable zone, the relationship between the transmission and the scanning time is depicted in Fig.~\ref{fig:ev}.
It took the system approximately $0.2$ ms to evolve from the low-lying to the high-lying state; thus, we can estimate the approximate relaxation time $T_r\approx 0.2$ ms.

\end{document}